\documentclass[prd,preprint,superscriptaddress,showpacs,byrevtex]{revtex4}
\usepackage{bm}
\def\fsl#1{\setbox0=\hbox{$#1$}           % set a box for #1
   \dimen0=\wd0                                 % and get its size
   \setbox1=\hbox{/} \dimen1=\wd1               % get size of /
   \ifdim\dimen0>\dimen1                        % #1 is bigger
      \rlap{\hbox to \dimen0{\hfil/\hfil}}      % so center / in box
      #1                                        % and print #1
   \else                                        % / is bigger
      \rlap{\hbox to \dimen1{\hfil$#1$\hfil}}   % so center #1
      /                                         % and print /
   \fi}                                         %
\newcommand{\be}{\begin{equation}}
\newcommand{\ee}{\end{equation}}
\newcommand{\bea}{\begin{eqnarray}}
\newcommand{\eea}{\end{eqnarray}}
\newcommand{\beq}{\begin{equation}}
\newcommand{\eeq}{\end{equation}}
\newcommand{\beqs}{\begin{eqnarray}}
\newcommand{\eeqs}{\end{eqnarray}}

\newcommand{\aslash}{A\hspace{-0.067in}\slash}

\begin{document}
\title{ Lattice QCD Method To Study Proton Radius Puzzle }
\author{Gouranga C Nayak }\thanks{E-Mail: nayakg138@gmail.com}
%
%\affiliation{ C. N. Yang Institute for Theoretical Physics, Stony Brook University, Stony Brook NY, 11794-3840 USA}
%
\date{\today}
\begin{abstract}
Recently there has been disagreement between various experiments about the value of the proton radius which is known as the proton radius puzzle. Since the proton is not a point particle the charge radius of the proton depends on the charge distribution (the form factor) of the partons inside the proton. Since this form factor is a non-perturbative quantity in QCD it cannot be calculated by using the perturbative QCD (pQCD) method but it can be calculated by using the lattice QCD method. In this paper we formulate the lattice QCD method to study the charge radius of the proton. We derive the non-perturbative formula of the charge radius of the proton from the first principle in QCD which can be calculated by using the lattice QCD method.
\end{abstract}
\maketitle
\pagestyle{plain}

\pagenumbering{arabic}

\section{ Introduction }

By using the spectroscopic method involving the electronic hydrogen (the ordinary hydrogen atom consisting of proton and electron) the charge radius of the proton is measured to be 0.8768 $\times 10^{-15}$ meter. Similarly by using the electron-proton scattering method the charge radius of the proton is measured to be 0.8775 $\times 10^{-15}$ meter which is consistent with the spectroscopic method. The CODATA-2014 world average value of the charge radius of the proton by using the electrons, {\it i. e.}, by using the above two methods, is 0.8751 $\times 10^{-15}$ meter \cite{pr}.

However, the muonic hydrogen (hydrogen atom consisting of proton and muon) experiment in the year 2010 found that the charge radius of the proton is 0.84087 $\times 10^{-15}$ meter \cite{mh,mh1}. The CODATA-2018 value of the charge radius of the proton is 0.8414 $\times 10^{-15}$ meter \cite{cod}. This disagreement between various experiments about the value of the charge radius of the proton is known as the proton radius puzzle which remains an unsolved problem in science.

It is well known that the proton is not a point particle but it is a composite particle consisting of quarks and gluons which are the fundamental particles of the nature. The up quark has the fractional electric charge $\frac{2e}{3}$ and the down quark has fractional electric charge $-\frac{e}{3}$ where $e$ is the magnitude of the charge of the electron. Hence the charge radius of the proton depends on the charge distribution (the form factor) of the partons inside the proton.

The electric charge radius $R_P$ of the proton $P$ is given by \cite{rp}
\bea
R^2_P=-\frac{6}{G_E(0)}\frac{dG_E(Q^2)}{dQ^2}|_{Q^2=0}
\label{rpj}
\eea
where $G_E(Q^2)$ is the electric form factor and $Q^2$ is the momentum transfer square of the virtual photon in the lepton-proton scattering.

The interaction between the quarks and gluons inside the proton is described by the quantum chromodynamics (QCD) \cite{ymj} which is a fundamental theory of the nature. The short distance partonic cross section can be calculated by using the perturbative QCD (pQCD) due to the asymptotic freedom in QCD \cite{gwj}. Using the factorization theorem in QCD \cite{fcj,fcj1,fcj2} the hadronic cross section can be calculated from the partonic cross section at the high energy colliders by using the experimentally extracted parton distribution function (PDF) and fragmentation function (FF).

The formation of the proton from the quarks and gluons is a long distance phenomenon in QCD. Due to the asymptotic freedom the QCD coupling becomes large at the large distance where the pQCD is not applicable. Hence the formation of the proton from the quarks and gluons cannot be studied by using the pQCD. The non-perturbative QCD is necessary to study the formation of the proton from the quarks and gluons.

However, the analytical solution of the non-perturbative QCD is not known yet because of the presence of the cubic and quartic power of the gluon fields in the QCD lagrangian inside the path integration in the generating functional in QCD [see section II for details]. The path integration in QCD can be performed numerically in the Euclidean time by using the lattice QCD method. Hence the lattice QCD provides the first principle method to study the formation of the proton from the quarks and gluons.

Since the electric form factor $G_E(Q^2)$ of the partons inside the proton in eq. (\ref{rpj}) is a non-perturbative quantity in QCD it cannot be calculated by using the perturbative QCD (pQCD) method but it can be calculated by using the lattice QCD method.

Recently we have formulated the lattice QCD method to study the proton formation from the quarks and gluons \cite{pj} and to study the proton spin crisis \cite{psj} and to study the proton decay \cite{pd} by implementing the non-zero boundary surface term in QCD which arises due to the confinement of quarks and gluons inside the finite size proton \cite{nkbsj}. In this paper we extend this to formulate the lattice QCD method to study the proton radius puzzle. We derive the non-perturbative formula of the charge radius of the proton from the first principle in QCD which can be calculated by using the lattice QCD method.

The paper is organized as follows. In section II we describe the lattice QCD method to study the proton formation from quarks and gluons by implementing the non-zero boundary surface term in QCD due to confinement. In section III we formulate the lattice QCD method to study the proton radius puzzle and derive the non-perturbative formula of the charge radius of the proton from the first principle in QCD which can be calculated by using the lattice QCD method. Section IV contains conclusions.

\section{Formation of proton from quarks and gluons Using lattice QCD Method}

The partonic operator for the proton ($P$) formation is given by
\bea
{\cal O}_P(x)=\epsilon_{kln} U_k^T(x) C\gamma_5 D_l(x) U_n(x)
\label{po}
\eea
where $U_k(x)$ is the up quark field and $D_k(x)$ is the down quark field, $C$ is the charge conjugation operator and $k,l,n=1,2,3$ are the color indices.

In the path integral formulation of QCD the vacuum expectation value of the non-perturbative partonic correlation function of the type $<0|{\cal O}^\dagger_P(x'){\cal O}_P(x'')|0>$ is given by
\bea
&&<0|{\cal O}^\dagger_P(x'){\cal O}_P(0)|0>=\frac{1}{Z[0]}\int [dA] [d{\bar U}][dU][d{\bar D}][dD] \times {\cal O}^\dagger_P(x'){\cal O}_P(0) \times {\rm det}[\frac{\delta B_f^s}{\delta \omega^b}]\nonumber \\
&& \times {\rm exp}[i\int d^4x [-\frac{1}{4} F_{\sigma \lambda}^s(x)F^{\sigma \delta s}(x) -\frac{1}{2\alpha} [B_f^s(x)]^2 +{\bar U}_k(x)[\delta^{kn}(i{\not \partial}-m_U)+gT^s_{kn}\aslash^s(x)]U_n(x)\nonumber \\
&&+{\bar D}_k(x)[\delta^{kn}(i{\not \partial}-m_D)+gT^s_{kn}\aslash^s(x)]D_n(x)]]
\label{pcf}
\eea
where $B_f^s(x)$ is the gauge fixing term with color index $s=1,...,8$, the $A_\sigma^s(x)$ is the gluon field with Lorentz index $\sigma=0,1,2,3$, the $\alpha$ is the gauge fixing parameter, $m_U$ is the mass of the up quark, $m_D$ is the mass of the down quark and $Z[0]$ is the generating functional in QCD given by
\bea
&&
Z[0]=\int [dA] [d{\bar U}][dU][d{\bar D}][dD] \times {\rm det}[\frac{\delta B_f^s}{\delta \omega^b}]\times {\rm exp}[i\int d^4x [-\frac{1}{4} F_{\sigma \lambda}^s(x)F^{\sigma \delta s}(x) -\frac{1}{2\alpha} [B_f^s(x)]^2 \nonumber \\
&&+{\bar U}_k(x)[\delta^{kn}(i{\not \partial}-m_U)+gT^s_{kn}\aslash^s(x)]U_n(x)+{\bar D}_k(x)[\delta^{kn}(i{\not \partial}-m_D)+gT^s_{kn}\aslash^s(x)]D_n(x)]]\nonumber \\
\label{pz0}
\eea
with
\bea
F_{\sigma \lambda}^s(x)=\partial_\sigma A_\lambda^s(x) - \partial_\lambda A_\sigma^s(x) +gf^{scd} A_\sigma^c(x) A_\lambda^d(x).
\label{fsl}
\eea
In eq. (\ref{pcf}) we do not have ghost fields as we directly work with the ghost determinant ${\rm det}[\frac{\delta B_f^s}{\delta \omega^b}]$.

The time evolution of the partonic operator in the Heisenberg representation is given by
\bea
{\cal O}_P(t,{\vec r})=e^{-iHt} {\cal O}_P(0,{\vec r}) e^{iHt}
\label{tv}
\eea
where the QCD Hamiltonian of the partons is given by $H$.

The complete set of proton energy-momentum eigenstates is given by
\bea
\sum_{l''} |H_{l''}><H_{l''}|=1.
\label{cse}
\eea

Using eqs. (\ref{tv}) and (\ref{cse}) in (\ref{pcf}) we find in the Euclidean time
\bea
&&\sum_{{\vec r}} <0|{\cal O}^\dagger_P(t,{\vec r}){\cal O}_P(0)|0>= \sum_{l''} |<H_{l''}|{\cal O}_P(0)|0>|^2~ e^{-\int dt~ E_{l''}(t)}
\label{pcfa}
\eea
where $\int dt$ is an indefinite integration. In the large time limit we neglect the higher energy level contribution to find
\bea
&&[\sum_{{\vec r}} <0|{\cal O}^\dagger_P(t,{\vec r}){\cal O}_P(0)|0>]_{t\rightarrow \infty} = |<P|{\cal O}_P(0)|0>|^2~ e^{-\int dt ~E(t)}
\label{pcfb}
\eea
where $E(t)$ is the energy of all the partons inside the proton in its ground state and $|P>$ is the energy-momentum eigenstate of the proton in its ground state.

Due to non-vanishing boundary surface term $E_{BS}(t)$ in QCD due to the confinement of partons inside the finite size proton we find \cite{nkbsj}
\bea
E_P=E(t)+E_{BS}(t)
\label{bs}
\eea
where $E_P$ is the energy of the proton, $E(t)$ is energy of all the partons inside the proton and $E_{BS}(t)$ is the non-vanishing boundary surface term in QCD
given by
\bea
\frac{dE_{BS}(t')}{dt'}=[\frac{\sum_{{\vec r}''} <0|{\cal O}^\dagger_P(t'',{\vec r}'')[\sum_{q,{\bar q},g}\int d^3r' \partial_i T^{i0}(t',{\vec r}')] {\cal O}_P(0)|0>}{\sum_{{\vec r}''} <0|{\cal O}^\dagger_P(t'',{\vec r}''){\cal O}_P(0)|0>}]_{t'' \rightarrow \infty}.
\label{bst}
\eea
In eq. (\ref{bst}) the energy-momentum tensor density $T^{\sigma \lambda}(x)$ of the partons in QCD is given by
\bea
&& T^{\sigma \lambda}(x) =F^{\sigma \mu s}(x)F_{\mu}^{~\lambda s}(x) +\frac{g^{\sigma \lambda}}{4} F^{\sigma' \mu' s}(x)F_{\mu'}^{~\lambda' s}(x) ++ {\bar U}_l(x) \gamma^\sigma [\delta^{ln}i\partial^\lambda -igT^s_{ln}A^{\lambda s}(x)]U_n(x)\nonumber \\
&& + {\bar D}_l(x) \gamma^\sigma [\delta^{ln}i\partial^\lambda -igT^s_{ln}A^{\lambda s}(x)]D_n(x).
\label{enmf}
\eea
Using eqs. (\ref{bs}) and (\ref{bst}) in (\ref{pcfb}) we find
\bea
&& |<P|{\cal O}_P(0)|0>|^2e^{- M_Pt}=[\frac{\sum_{{\vec r}'} <0|{\cal O}^\dagger_P(t',{\vec r}') {\cal O}_P(0)|0>}{e^{\int dt' [\frac{\sum_{{\vec r}''} <0|{\cal O}^\dagger_P({\vec r}'',t'')[\sum_{q,{\bar q},g}\int dt' \int d^3r' \partial_i T^{i0}(t',{\vec r}') ]{\cal O}_P(0)|0>}{\sum_{{\vec r}''} <0|{\cal O}^\dagger_P({\vec r}'',t''){\cal O}_P(0)|0>}]_{t'' \rightarrow \infty}}}]_{t' \rightarrow \infty}\nonumber \\
\label{frs}
\eea
which can be calculated by using the lattice QCD method where $M_P$ is the mass of the proton and $\int dt'$ is indefinite integration.

Eq. (\ref{frs}) is the non-perturbative formula to study the proton formation from quarks and gluons using the lattice QCD method by implementing non-zero boundary surface term in QCD due to confinement.

\section{Lattice QCD Method to study charge radius of the proton }

In the previous section we have formulated the lattice QCD method to study the formation of the proton from the quarks and gluons by implementing the non-zero boundary surface term in QCD due to the confinement. In this section we will extend this to formulate the lattice QCD method to study the proton radius puzzle.
We will derive the non-perturbative formula of the charge radius of the proton from the first principle in QCD which can be calculated by using the lattice QCD method. 

This method is also used to study various quantities in QCD in vacuum \cite{psj,pd,avj} and in QCD in medium  \cite{amj} to study the quark-gluon plasma at RHIC and LHC \cite{qk,qk1,qk2}.

In the single (virtual) photon exchange the amplitude ${\cal M}$ for the electron-proton ($eP$) elastic scattering process is given by
\bea
{\cal M} =\frac{g^{\lambda \delta}}{q^2} [{\bar u}_e(k_f) \gamma_\lambda u_e(k_i)][ie{\bar u}_P(p_f) \Gamma_\delta(p_f,p_i)u_P(p_i)]  =\frac{4\pi \alpha}{Q^2} l_{\lambda} J^\lambda
\label{scm}
\eea
where $k_i (k_f)$, $p_i(p_f)$ are the initial (final) momentum of the electron, proton, $q^\mu=p^\mu_f-p^\mu_i$ is the momentum of the virtual photon, $u_e,u_P$ are the electron, proton spinors respectively, $l^\mu,J^\mu$ are the leptonic, hadronic (electromagnetic) currents respectively and $q^2=-Q^2$. The general form of the vertex function $\Gamma^\lambda(p_f,p_i)$ satisfying the relativistic invariance and current conservation is given by
\bea
J^\lambda = {\bar u}_P(p_f) [\gamma^\lambda F_1(Q^2)+\frac{\sigma^{\lambda \delta}q_\delta}{2M_P}F_2(Q^2) ]u_P(p_i)=<P|\frac{2}{3} {\bar U} \gamma^\lambda U - \frac{1}{3} {\bar D} \gamma^\lambda D|P>.
\label{hdc}
\eea
The electric form factor $G_E(Q^2)$ is given by \cite{rp}
\bea
G_E(Q^2)=F_1(Q^2)-\frac{Q^2}{4M_P^2}F_2(Q^2).
\label{geq}
\eea

Using eqs. (\ref{tv}) and (\ref{cse}) in (\ref{pcf}) we find in the Euclidean time
\bea
&&\sum_{{\vec r}} e^{{\vec p}\cdot {\vec r}} <0|{\cal O}^\dagger_P(t,{\vec r}){\cal O}_P(0)|0>= \sum_{l''} |<H_{l''}({\vec p})|{\cal O}_P(0)|0>|^2~ e^{-\int dt~ E_{l''}({\vec p},t)}
\label{pcfa3}
\eea
where ${\vec p}$ is the momentum of the proton and $\int dt$ is an indefinite integration. In the large time limit we neglect the higher energy level contribution in eq. (\ref{pcfa3}) to find
\bea
&&[\sum_{{\vec r}} e^{{\vec p}\cdot {\vec r}} <0|{\cal O}^\dagger_P(t,{\vec r}){\cal O}_P(0)|0>]_{t\rightarrow \infty} = |<P({\vec p})|{\cal O}_P(0)|0>|^2~ e^{-\int dt ~E({\vec p},t)}
\label{pcfb3}
\eea
where $E({\vec p},t)$ is the energy of all the partons inside the proton of momentum ${\vec p}$ in its ground state and $|P({\vec p})>$ is the energy-momentum eigenstate of the proton $P$ of momentum ${\vec p}$ in its ground state.

From eq. (\ref{hdc}) the electromagnetic current operator $j^\lambda_q(x)$ of the quarks inside the proton is given by
\bea
j_\lambda^q(x) = \frac{2}{3} {\bar U}(x) \gamma_\lambda U(x) - \frac{1}{3} {\bar D}(x) \gamma_\lambda D(x)=\sum_f e_f {\bar q}_f(x)\gamma_\lambda q(x)
\label{qem}
\eea
where $q_f(x)$ and $e_f$ are the quark field and the fractional electrical charge of the quark of the flavor $f$ respectively.

In the path integral formulation of QCD the vacuum expectation value of the non-perturbative 3-point partonic correlation function of the type $<0|{\cal O}_P(x'')j^q_0(x'){\cal O}_P(0)|0>$ is given by
\bea
&&<0|{\cal O}^\dagger_P(x'')j^q_0(x'){\cal O}_P(0)|0>=\frac{1}{Z[0]}\int [dA] [d{\bar U}][dU][d{\bar D}][dD] \times {\cal O}^\dagger_P(x'')j^q_0(x'){\cal O}_P(0) \times {\rm det}[\frac{\delta B_f^s}{\delta \omega^b}]\nonumber \\
&& \times {\rm exp}[i\int d^4x [-\frac{1}{4} F_{\sigma \lambda}^s(x)F^{\sigma \delta s}(x) -\frac{1}{2\alpha} [B_f^s(x)]^2 +{\bar U}_k(x)[\delta^{kn}(i{\not \partial}-m_U)+gT^s_{kn}\aslash^s(x)]U_n(x)\nonumber \\
&&+{\bar D}_k(x)[\delta^{kn}(i{\not \partial}-m_D)+gT^s_{kn}\aslash^s(x)]D_n(x)]].
\label{pcf3}
\eea
In this paper we assume that the initial proton is at rest, {\it i. e.}, ${\vec p}_i=0$. This means ${\vec p}_f={\vec q}={\vec p}$. Using eqs. (\ref{tv}) and (\ref{cse}) in (\ref{pcf3}) we find in the Euclidean time
\bea
&& \sum_{{\vec r}'',{\vec r}'} e^{i{\vec q}\cdot ({\vec r}''-{\vec r}')} <0|{\cal O}_P(t'',{\vec r}'')j^q_0(t',{\vec r}') {\cal O}_P(0)|0> =\sum_{n'',n'}<0|{\cal O}_P|H_{n''}({\vec p})>\nonumber \\
&&<H_{n''}({\vec p})|j^q_0|H_{n'}><H_{n'}|{\cal O}_P|0>e^{-[\int dt'' E_{n''}({\vec p},t'')-\int dt' E_{n''}({\vec p},t')]}e^{-\int dt' E_{n'}(t')}
\label{mef}
\eea
where ${\vec p}$ is the momentum of the final state proton and ${\vec q}={\vec p}$.

In the limit $t'' >>>t',~~t'\rightarrow \infty$ we find by neglecting the higher energy level contributions
\bea
&& [\sum_{{\vec r}'',{\vec r}'} e^{i{\vec q}\cdot ({\vec r}''-{\vec r}')} <0|{\cal O}_P(t'',{\vec r}'')j^q_0(t',{\vec r}') {\cal O}_P(0)|0>]_{t''>>>t',~~t'\rightarrow \infty} =<0|{\cal O}_P|P({\vec p})>\nonumber \\
&&<P({\vec p})|j^q_0|P><P|{\cal O}_P|0>e^{-[\int dt'' E({\vec p},t'')-\int dt' E({\vec p},t')]}~e^{-\int dt' E(t')}
\label{meg3}
\eea
where $E({\vec p},t)$ is the energy of all the partons inside the proton of momentum ${\vec p}$, the $E(t)$ is the energy of all the partons inside the proton at rest,
$|P({\vec p})>$ is the energy-momentum eigenstate of the proton of momentum ${\vec p}$ and $|P>$ is the energy-momentum eigenstate of the proton at rest.

From eq. (\ref{pcfb}) we find for the proton at rest
\bea
&& [\sum_{{\vec r}'} <0|{\cal O}_P(t',{\vec r}') {\cal O}_P(0)|0>]_{t' \rightarrow \infty} =|<P|{\cal O}_P|0>|^2e^{-\int dt' E(t')}.
\label{meh3}
\eea
Similarly from eq. (\ref{pcfb3}) we find for the proton with momentum ${\vec p}$
\bea
&& [\sum_{{\vec r}''} e^{i{\vec p}\cdot {\vec r}''} <0|{\cal O}_P({\vec r}'',t''-t') {\cal O}_P(0)|0>]_{t''>>>t',~~t' \rightarrow \infty} \nonumber \\
&&=|<P({\vec p})|{\cal O}_P|0>|^2e^{-[\int dt'' E({\vec p},t'')-\int dt' E({\vec p},t')]}.
\label{mei3}
\eea
From eqs. (\ref{meg3}), (\ref{meh3}) and (\ref{mei3}) we find
\bea
&&<P({\vec p})|j^q_0|P>= \sqrt{|<P({\vec p})|{\cal O}_P|0>|^2}\times \sqrt{|<P|{\cal O}_P|0>|^2}\nonumber \\
&&\times [\frac{\sum_{{\vec r}''',{\vec r}'} e^{i{\vec q}\cdot ({\vec r}'-{\vec r}''')} <0|{\cal O}_P(t',{\vec r}')j^q_0(t''',{\vec r}''') {\cal O}_P(0)|0>}{[\sum_{{\vec r}'''} <0|{\cal O}_P(t''',{\vec r}''') {\cal O}_P(0)|0>][\sum_{{\vec r}'} e^{i{\vec p}\cdot {\vec r}'} <0|{\cal O}_P(t'-t''',{\vec r}') {\cal O}_P(0)|0>]}]_{t'>>>t''',~~~t'''\rightarrow \infty}.\nonumber \\
\label{mej3}
\eea

For the initial proton at rest with energy-momentum eigenstate $|P>$ and the final proton of momentum ${\vec p}$ with energy-momentum eigenstate $|P({\vec p})>$ we find that the electric form factor $G_E(Q^2)$ of the partons inside the proton is related to the proton matrix element $<P({\vec p})|j^q_0|P>$ of the quark electromagnetic current $j^q_0$ via the relation \cite{eff}
\bea
G_E(Q^2)=\sqrt{\frac{2E_P({\vec p})}{E_P({\vec p})+M_P}}<P({\vec p})|j^q_0|P>
\label{mpge}
\eea
where $E_P({\vec p})$ is the energy of the proton of momentum ${\vec p}={\vec q}$ and $Q^2=-q^2$.

From eq. (\ref{frs}) we find for the proton at rest
\bea
&& |<P|{\cal O}_P(0)|0>|^2=[\frac{\sum_{{\vec r}''} <0|{\cal O}^\dagger_P(t'',{\vec r}'') {\cal O}_P(0)|0>\times e^{ M_Pt''}}{e^{\int dt'' [\frac{\sum_{{\vec r}'''} <0|{\cal O}^\dagger_P(t''',{\vec r}''')[\sum_{q,{\bar q},g}\int dt'' \int d^3r'' \partial_i T^{i0}(t'',{\vec r}'')] {\cal O}_P(0)|0>}{\sum_{{\vec r}'''} <0|{\cal O}^\dagger_P(t''',{\vec r}'''){\cal O}_P(0)|0>}]_{t''' \rightarrow \infty}}}]_{t'' \rightarrow \infty}\nonumber \\
\label{mek3}
\eea
where $M_P$ is the mass of the proton. Similarly for the proton with momentum ${\vec p}$ we find
\bea
&& |<P({\vec p})|{\cal O}_P(0)|0>|^2=[\frac{\sum_{{\vec r}''} e^{i{\vec p}\cdot {\vec r}''}<0|{\cal O}^\dagger_P(t'',{\vec r}'') {\cal O}_P(0)|0>\times e^{t'' E_P({\vec p})}}{e^{\int dt'' [\frac{\sum_{{\vec r}'''} e^{i{\vec p}\cdot {\vec r}'''}<0|{\cal O}^\dagger_P(t''',{\vec r}''')[\sum_{q,{\bar q},g}\int dt'' \int d^3r'' \partial_i T^{i0}(t'',{\vec r}'')] {\cal O}_P(0)|0>}{\sum_{{\vec r}'''} e^{i{\vec p}\cdot {\vec r}'''}<0|{\cal O}^\dagger_P(t''',{\vec r}'''){\cal O}_P(0)|0>}]_{t''' \rightarrow \infty}}}]_{t'' \rightarrow \infty}\nonumber \\
\label{mel3}
\eea
where $E_P({\vec p})$ is the energy of the proton with momentum ${\vec p}$.

Using eqs. (\ref{mek3}), (\ref{mel3}) and (\ref{mej3}) in (\ref{mpge}) we find
\bea
&&G_E(Q^2)=\sqrt{\frac{2E_P({\vec p})}{E_P({\vec p})+M_P}}\nonumber \\
&& \times \left[[\frac{\sum_{{\vec r}''} e^{i{\vec p}\cdot {\vec r}''}<0|{\cal O}^\dagger_P(t'',{\vec r}'') {\cal O}_P(0)|0>\times e^{t'' E_P({\vec p})}}{e^{\int dt'' [\frac{\sum_{{\vec r}'''} e^{i{\vec p}\cdot {\vec r}'''}<0|{\cal O}^\dagger_P(t''',{\vec r}''')[\sum_{q,{\bar q},g}\int dt'' \int d^3r'' \partial_i T^{i0}(t'',{\vec r}'')] {\cal O}_P(0)|0>}{\sum_{{\vec r}'''} e^{i{\vec p}\cdot {\vec r}'''}<0|{\cal O}^\dagger_P(t''',{\vec r}'''){\cal O}_P(0)|0>}]_{t''' \rightarrow \infty}}}]_{t'' \rightarrow \infty}\right]^{\frac{1}{2}}\nonumber \\
&& \times \left[[\frac{\sum_{{\vec r}''} <0|{\cal O}^\dagger_P(t'',{\vec r}'') {\cal O}_P(0)|0>\times e^{ M_Pt''}}{e^{\int dt'' [\frac{\sum_{{\vec r}'''} <0|{\cal O}^\dagger_P(t''',{\vec r}''')[\sum_{q,{\bar q},g}\int dt'' \int d^3r'' \partial_i T^{i0}(t'',{\vec r}'')] {\cal O}_P(0)|0>}{\sum_{{\vec r}'''} <0|{\cal O}^\dagger_P(t''',{\vec r}'''){\cal O}_P(0)|0>}]_{t''' \rightarrow \infty}}}]_{t'' \rightarrow \infty} \right]^{\frac{1}{2}} \nonumber \\
&&\times [\frac{\sum_{{\vec r}''',{\vec r}'} e^{i{\vec q}\cdot ({\vec r}'-{\vec r}''')} <0|{\cal O}_P(t',{\vec r}')j^q_0(t''',{\vec r}''') {\cal O}_P(0)|0>}{[\sum_{{\vec r}'''} <0|{\cal O}_P(t''',{\vec r}''') {\cal O}_P(0)|0>][\sum_{{\vec r}'} e^{i{\vec p}\cdot {\vec r}'} <0|{\cal O}_P(t'-t''',{\vec r}') {\cal O}_P(0)|0>]}]_{t'>>>t''',~~~t'''\rightarrow \infty}\nonumber \\
\label{mem3}
\eea
where the initial proton is at rest and the final proton momentum is ${\vec p}={\vec q}$ with $Q^2=-q^2$.

The eq. (\ref{mem3}) is the non-perturbative formula of the electric form factor $G_E(Q^2)$ of the proton derived from the first principle in QCD which can be calculated by using the lattice QCD method. The charge radius $R_P$ of the proton is obtained from this electric form factor $G_E(Q^2)$ in eq. (\ref{mem3}) by using eq. (\ref{rpj}).

\section{Conclusions}
Recently there has been disagreement between various experiments about the value of the proton radius which is known as the proton radius puzzle. Since the proton is not a point particle the charge radius of the proton depends on the charge distribution (the form factor) of the partons inside the proton. Since this form factor is a non-perturbative quantity in QCD it cannot be calculated by using the perturbative QCD (pQCD) method but it can be calculated by using the lattice QCD method. In this paper we have formulated the lattice QCD method to study the charge radius of the proton. We have derived the non-perturbative formula of the charge radius of the proton from the first principle in QCD which can be calculated by using the lattice QCD method.

\end{document}